\documentstyle[aps,preprint,prb,epsfig,rotating,amsmath,amssymb]{revtex}
\draft
\title{Influence of shape of quantum dots on their\\
       far-infrared absorption}
\author{Ingibj{\"o}rg Magn{\'u}sd{\'o}ttir and Vidar Gudmundsson} 
\address{Science Institute, University of Iceland, 
         Dunhaga 3, IS-107 Reykjavik, Iceland}
\date{June 22, 1999}
\begin{document}
\tighten
\maketitle
%

\begin{abstract} 
We investigate the effects of the shape of quantum dots
on their far-infrared absorption in an external 
magnetic field by a model calculation. We focus our attention on
dots with a parabolic confinement potential deviating from the common
circular symmetry, and dots having circular doughnut shape. 
For a confinement where the generalized Kohn theorem does not hold
we are able to interprete the results in terms of a mixture
of a center-of-mass mode and collective modes reflecting
an excitation of relative motion of the electrons.   
The calculations are performed within the time-dependent
Hartree approximation and the results are compared to 
available experimental results.
\end{abstract}


\narrowtext
\section{Introduction}

Ever since the discovery that far-infrared radiation (FIR)
can only be used to excite center-of-mass modes of
electrons parabolically confined in circular quantum
dots, the generalized Kohn 
theorem,\cite{Maksym90:108,Brey89:10647,Bakshi90:7416}  
ingenious ways have been thought of to modify the
confinement in order to excite internal modes. 
The modes caused by relative motion of the electrons
would allow for exploration of interaction and general
many-body effects.  
It was demonstrated that in circular dots with a slight radial deviation 
from the parabolic confinement states with few electrons have
a unique far-infrared spectrum.\cite{Pfannkuche94:1221}
Dots with a higher number of electrons ($N>10$) commonly 
only show a simpler structure of Bernstein 
modes\cite{Pfannkuche94:1221,Gudmundsson95:17744}
and an energy shift.\cite{Gudmundsson91:12098,Ye94:17217} 
Early on, it came clear that a certain anticrossing behavior 
discovered in the higher one of the 
two Kohn's modes measured\cite{Demel90:788a}
for quantum dots assumed to be circular is indeed a signature
of a slight square deformation of the electron 
system.\cite{Pfannkuche91:13132,Gudmundsson91:12098}
The far-infrared absorption 
of strictly square-shaped quantum dots with hard walls has
been modelled by exact numerical diagonalization for two
electrons,\cite{Ugajin95:10714,Ugajin96:6963} 
and by a real-time simulation of the density oscillations
described by a local spin-density approximation 
(LSDA) for many electrons.\cite{Puente99:05288} 
The generalized Kohn theorem has been further extended to describe 
elliptic\cite{Yip91:1707} quantum dots and related 
three-dimensional structures, ellipsoids.\cite{Li91:5151}

The production of arrays of isolated angularly deformed quantum
dots has been hampered by technical difficulties.
The little there is of experimental results from FIR absorption
measurements on quantum dots with angular deviation
it has not found its way into the general physics literature
but is to be found as thesis 
work.\cite{Hochgrafe98:thesis,Vasiliadou95:thesis,Dahl92:thesis} 

The FIR absorption of single quantum rings has been studied in relatively
large systems with many electrons,\cite{Zaremba96:R10512,Emperador99:xx}
but recently small quantum dots with few electrons and a hole through
their center have been produced and measured.\cite{Lorke98:424} 

In this publication we model the FIR absorption of single quantum dots
with a slight elliptic or square deviation assuming the radial
confinement to be parabolic. The model is general enough to be
applicable for most angular deviations of the confinement.  
In a circular quantum dot the electron-electron interaction does
not break the angular symmetry, but in our model it can strongly
modify the shape of the dot as soon as the circular symmetry is
broken by the initial confinement. We apply our model to the
FIR absorption of an elliptic dot in order to verify that it
produces the results expected by Kohn's theorem.\cite{Yip91:1707}
The calculated absorption spectra for a square deviated quantum dot
show both familiar results for a weak 
deviation,\cite{Demel90:788a,Pfannkuche91:13132} 
and effects reflecting internal electron motion that have not yet
been observed in experiments. 

In addition, we calculate the FIR absorption for a small dot 
with few electrons with its center removed to compare with the
results of A.\ Lorke.\cite{Lorke98:424}

\section{Model}
\begin{figure} 
\epsfxsize 12cm 
\begin{center}
\epsffile{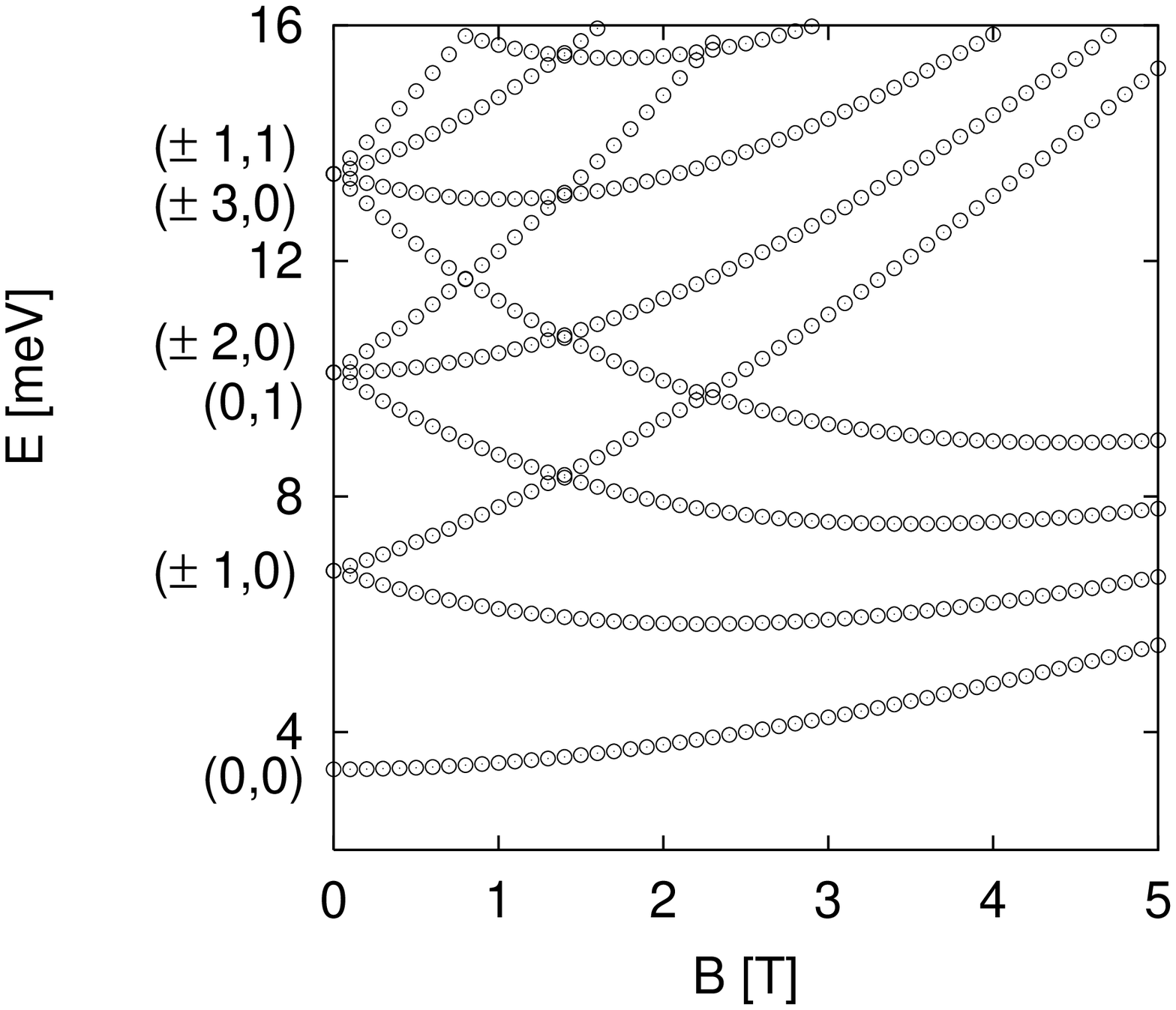}
\end{center}
\caption{The Darwin-Fock energy diagram of a single electron in a
circular parabolic quantum dot. At $B=0$, we have the familiar
two-dimensional harmonic oscillator. The states at $B=0$ are labelled
with $(M,n_r)$. }
\label{E_IM_df_0}
\end{figure}

As we allow for a very general angular shape of the quantum dot 
neither the total angular momentum nor the angular momentum of the
effective single particle states in a mean field approach is
conserved. The Coulomb interaction thus 'mixes` up all elements
of the functional basis used and we limit ourselves to the 
Hartree approximation (HA) in order to be able to calculate
the absorption. 
The quantum dot is modelled with the confinement potential
\begin{equation}
      V_{\text{conf}}({\bf r})=\frac{1}{2}m^*\omega_0^2r^2
      \Big[1+\sum_{p=1}^{p_{max}}\alpha_p\cos(2p\phi)\Big],
\end{equation}
representing an elliptic confinement when $\alpha_1\neq 0$ and
$\alpha_p=0$ for $p\neq 1$, and a square symmetric confinement when
$\alpha_2\neq 0$ and $\alpha_p=0$ for $p\neq 2$. We use the Darwin-Fock
basis; the eigenfunctions of the circular parabolic confinement
potential in which the natural length scale, $a$, is given by
\begin{equation}
a^2=\frac{\ell ^2}{\sqrt{1+4(\frac{\omega_0}{\omega_c})^2}},
\quad \ell ^2=\frac{\hbar c}{eB},
\end{equation}
where $\omega_c=eB/m^*c$ is the cyclotron frequency of an electron with
effective mass $m^*$ in a perpendicular homogeneous magnetic field
$B$. The states are labelled by the radial quantum number
$n_r$ and the angular quantum number $M$.\cite{Gudmundsson91:12098} 
The single electron
spectrum of these states is shown in Fig.\ \ref{E_IM_df_0}.   

To evaluate the FIR response of the
quantum dot, the external potential is assumed to have the form
\begin{equation}
      \phi^{ext}({\bf r},t)=\phi_0re^{i(N_p\phi-(\omega+i\eta)t)},
\end{equation}
where $N_p=\pm 1$ and $\eta\rightarrow 0^+$, representing a spatially
constant external electric field. The FIR response is found by a
self-consistent method in the linear response regime; the
time-dependent Hartree approximation. The power absorption is then
given by
\begin{equation}
      {\mathcal P}(\omega)\varpropto \omega\Im \Big\{\sum_{\alpha,\beta}
      f^{\beta\alpha}(\omega)\Big|\langle\beta|\phi^{sc}
      |\alpha\rangle\Big|^2\Big\},
\end{equation}
where
\begin{equation}
      f^{\beta\alpha}(\omega)=\frac{1}{\hbar}\Big\{\frac{f_\alpha^0-f_\beta^0}
      {\omega+(\omega_\alpha-\omega_\beta)+i\eta}\Big\}
\end{equation}
and $f^0$ is the equilibrium Fermi distribution. Self-consistency is
obtained by calculating linear response not to the external perturbation,
$\phi^{ext}$, but rather to the total (self-consistent) potential
$\phi^{sc}=\phi^{ext}+\phi_H^{ind}$, where $\phi_H^{ind}$ is the
induced Hartree potential.
For a circular quantum dot, the parabolic potential has
proven to be a realistic approximation in many cases. 
For such a circular
harmonic dot, the dipole selection rule for the
center-of-mass angular momentum is $\Delta M=N_p$ and the resonance
frequencies are given by
\begin{equation*}
      \omega_\pm=\frac{1}{2}(\Omega+N_p\omega_c),
      \quad \Omega=(\omega_c^2+4\omega_0^2)^{1/2}.
\end{equation*}
For an anisotropic harmonic confinement, the selection rule is still
$\Delta M=\pm 1$, but there is absorption into both $\omega_+$ and
$\omega_-$ for each polarization. The resonance frequencies are then
given by
\begin{equation}
      \omega_\pm^2=\frac{\omega_x^2+\omega_y^2+\omega_c^2
      \pm[\omega_c^4+2\omega_c^2(\omega_x^2+\omega_y^2)+
      (\omega_x^2-\omega_y^2)^2]^{1/2}}{2}.
\label{eq:reselliptic2}
\end{equation}
where $\omega_x$ and $\omega_y$ are the resonances at
$B=0$~T,~\cite{Li91:5151,Yip91:1707} fulfilling
$\omega_x=(1+\alpha_1)^{1/2}\omega_0$ and
$\omega_y=(1-\alpha_1)^{1/2}\omega_0$ in our model.
\section{Results}
\begin{figure} 
\epsfxsize 12cm 
\begin{center}
\epsffile{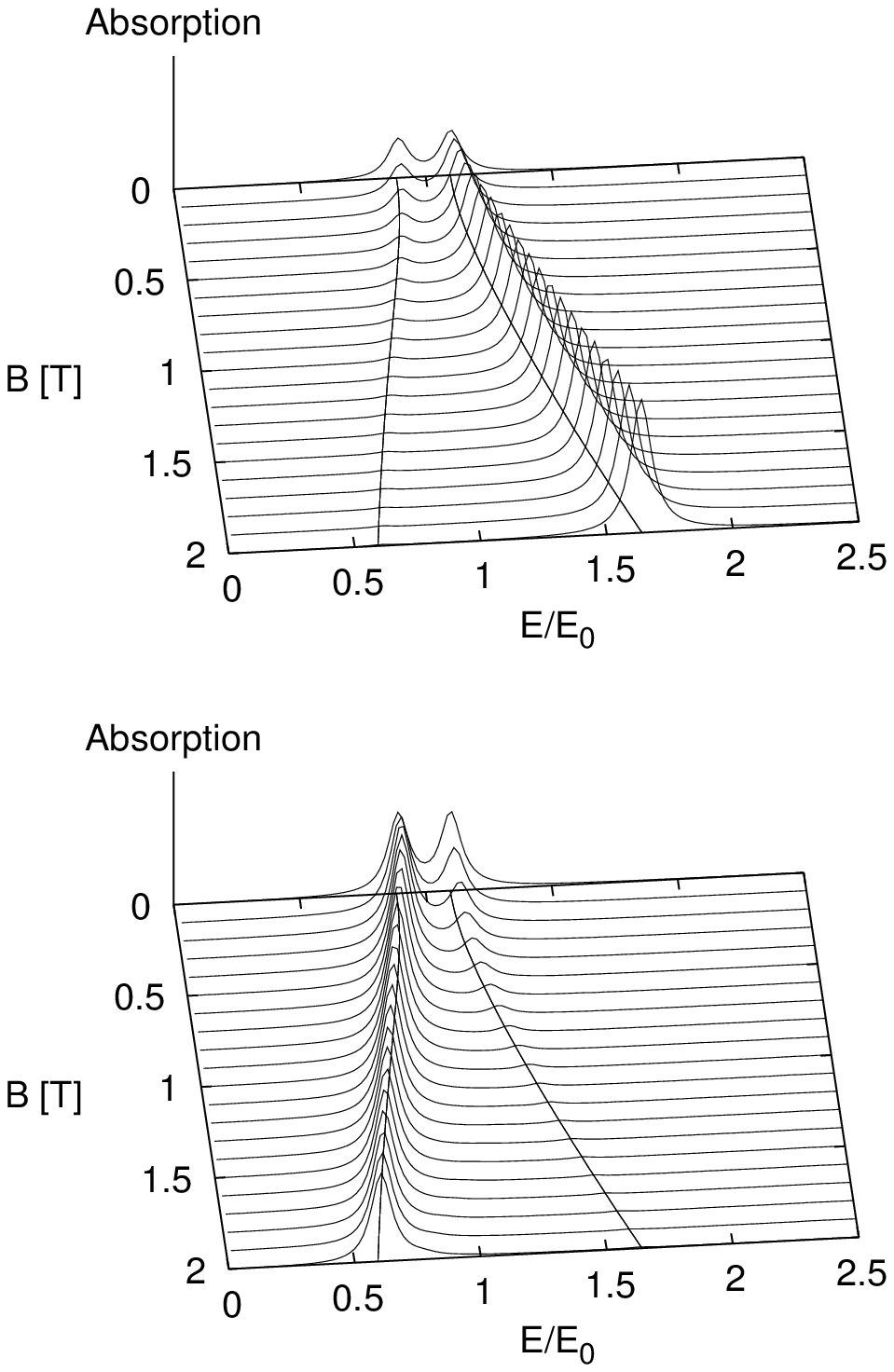}
\end{center}
\caption{The far-infrared absorption for $N_p=+1$ (above) and $N_p=-1$ 
(below) in the case of a two electron elliptic quantum dot
($\alpha_1=0.20$). The solid lines indicate the dispersion relation
for the absorption, with $\omega_x/\omega_0=1.10$ and $\omega_y/\omega_0=0.89$.}
\label{fig_P_IM_el}
\end{figure}

In the calculations reported here we use GaAs parameters, the
effective mass $m^*=0.067m_0$ and the dielectric constant $\kappa=12.4$.
The FIR absorption of a quantum dot with an elliptic deviation is
seen in Fig.\ \ref{fig_P_IM_el}. The induced density for the
oscillation modes was used to confirm that only center-of-mass
modes are excited and the dispersion with respect to $B$
can be reproduced as a simple difference of the dipole active
transfers in the Darwin-Fock energy spectrum for one electron 
in a parabolic elliptic confinement. The oscillation strengths 
do comply with known analytic expressions.\cite{Yip91:1707} 
Interestingly, the external circular polarized electric field
leads to linear oscillations of the electron system parallel
to the minor and major axis of the contours
of the elliptic confinement potential.

\begin{figure} 
\epsfxsize 12cm 
\begin{center}
\epsffile{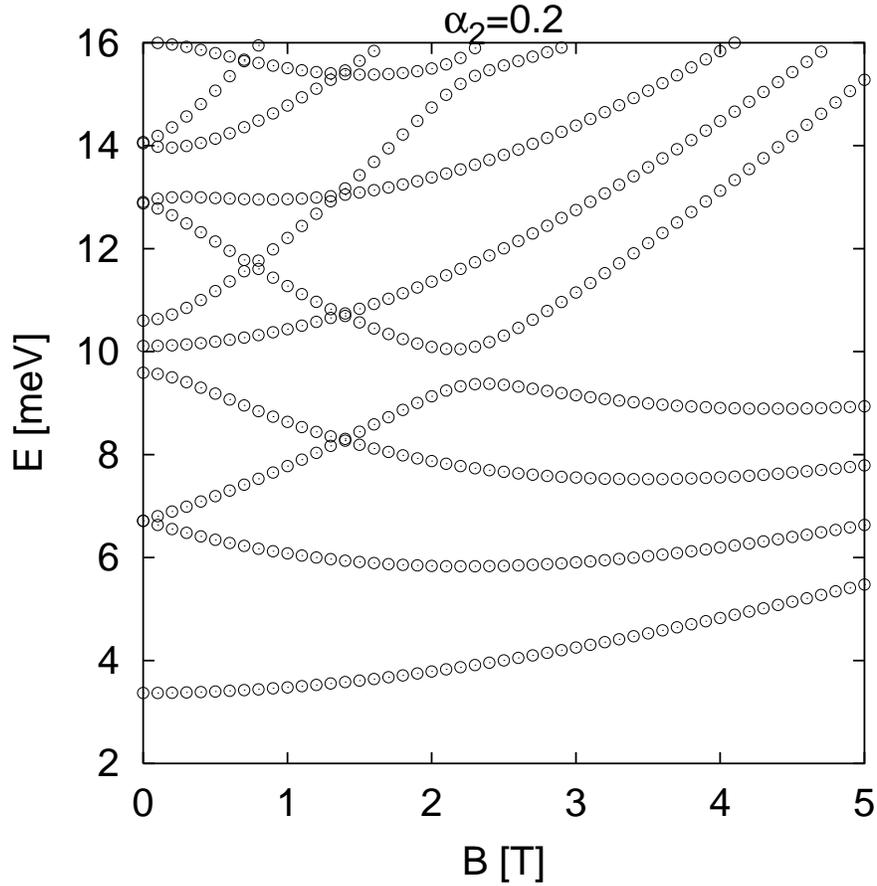}
\end{center}
\caption{The Darwin-Fock energy diagram of a single electron in a
square symmetric quantum dot with $\alpha_2=0.20$. The angular part of 
$V_\phi$, $\cos(4\phi)$, couples states in the circular parabolic
potential for which $\Delta M=\pm 4$. Degeneracy of states at $B=0$
fulfilling $\Delta M=\pm 4$ is lifted. An anticrossing occurs at
$B\approx 2.3T$ due to coupling of the accidentally degenerate states
$(+1,0)$ and $(-3,0)$ at this magnetic field.}
\label{E_IM_df_sq}
\end{figure}

\begin{figure} 
\epsfxsize 12cm 
\begin{center}
\epsffile{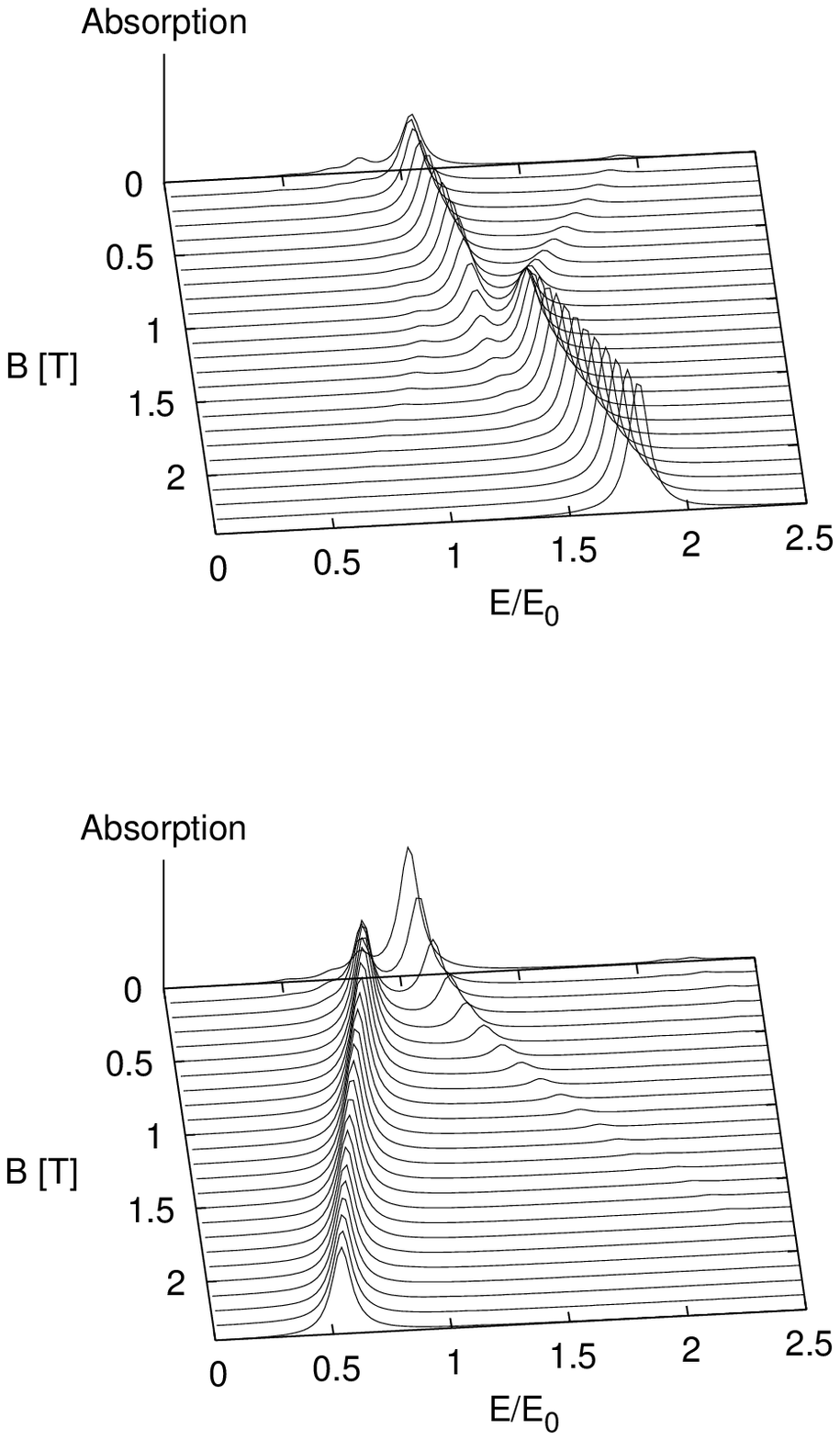}
\end{center}
\caption{The far-infrared absorption of a two electron square
symmetric quantum dot ($\alpha_2=0.20$) for $N_p=+1$ (above) and
$N_p=-1$ (below). $E_0=3.37$~meV, $T=1.0$~K.}
\label{fig_P_IM_sq1}
\end{figure}

\begin{figure} 
\epsfxsize 12cm 
\begin{center}
\epsffile{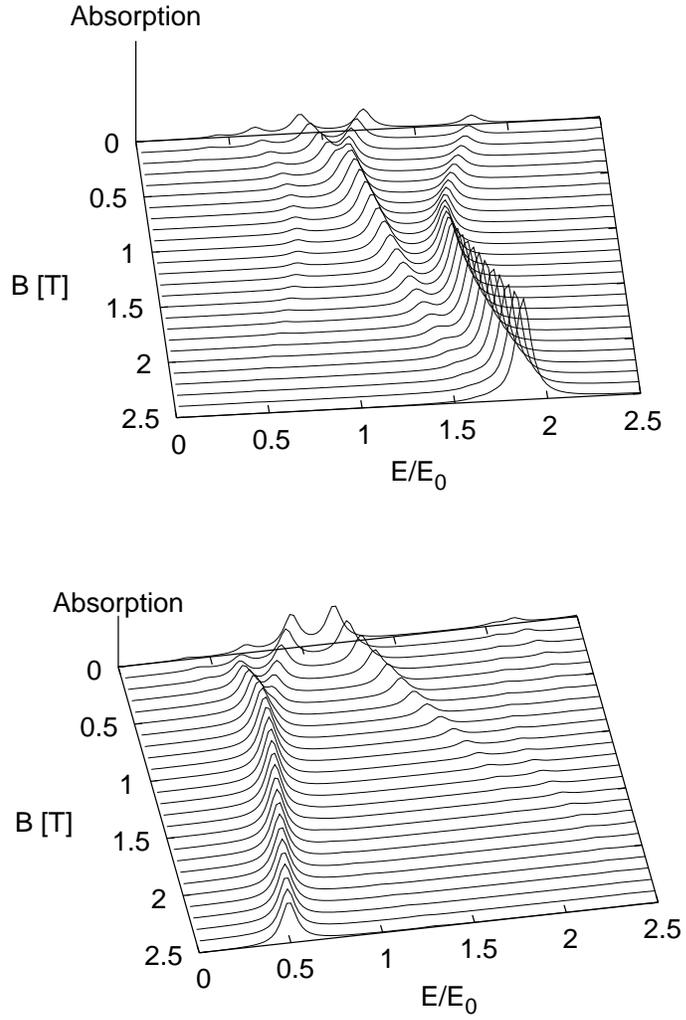}
\end{center}
\caption{The FIR absorption of a square symmetric two electron quantum
dot with $\alpha_2=0.40$, $N_p=+1$ (above) and $N_p=-1$
(below). $E_0=3.37$~meV, $T=1.0$~K. }
\label{fig_P_IM_sq2}
\end{figure}

\begin{figure} 
\epsfxsize 12cm 
\begin{center}
\epsffile{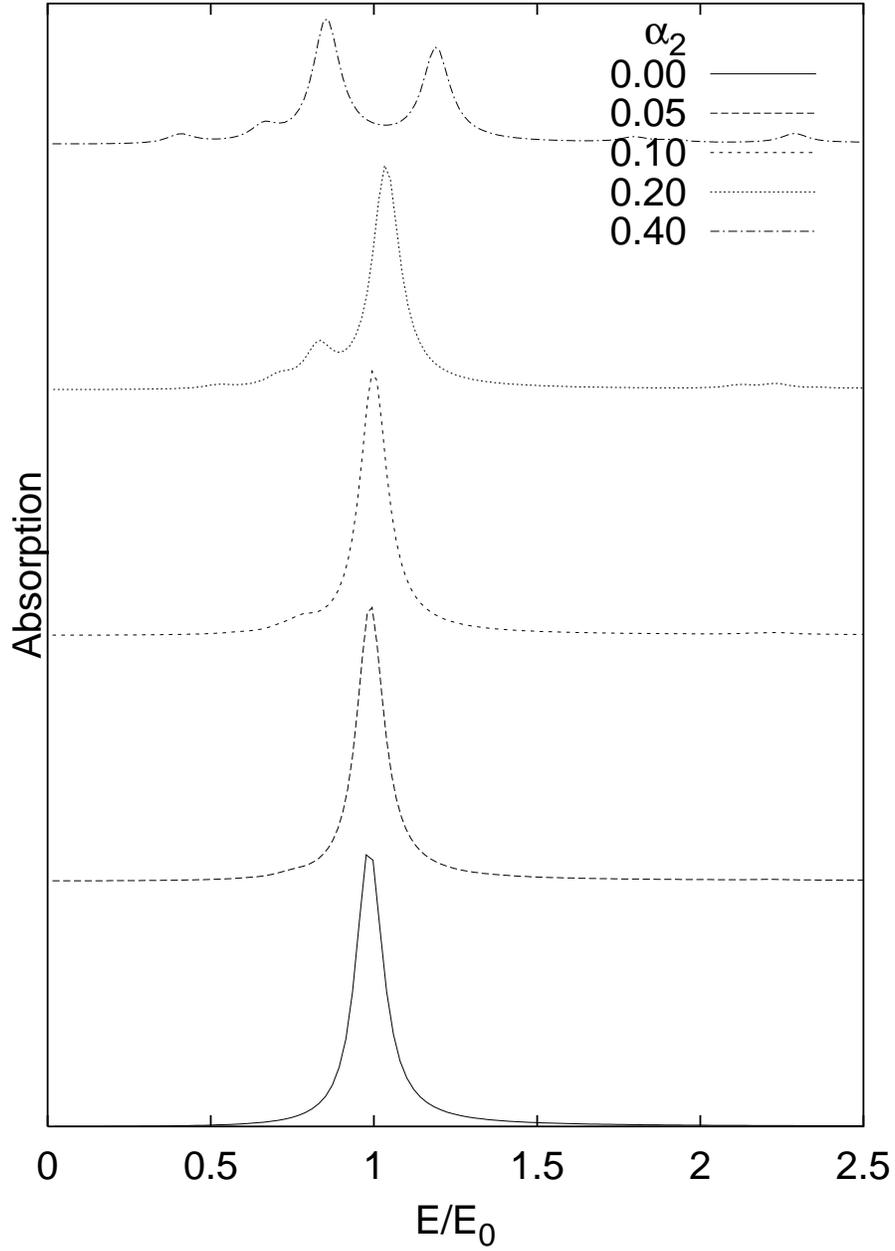}
\end{center}
\caption{The evolution of absorption peaks of a square symmetric two
electron quantum dot at $B=0T$ as $\alpha_2$ is increased. Influence
of the square symmetric deviation from the circular parabolic
confinement starts to show up in the absorption spectrum for
$\alpha_2\geq 0.10$. }
\label{fig_P_IM_sq2_ph}
\end{figure}

Figure \ref{E_IM_df_sq} shows the Darwin-Fock energy for a single
electron in a confining potential with square deviation. Clearly
visible is the anticrossing of the ($M$,$n_r$)$=$(+1,0) and the
state (-3,0) that was found to be visible in the FIR absorption
of quantum dots which were clearly of square shape in an electron 
micrograph.\cite{Demel90:788a,Pfannkuche91:13132,Gudmundsson91:12098}  
This anticrossing is present in the calculated absorption shown
in Fig.\ \ref{fig_P_IM_sq1} and \ref{fig_P_IM_sq2} for two electrons
in a dot. For this relative large deviation ($\alpha_2=0.2$ or $0.4$)
from circular symmetry the absorption at $B=0$ for both circular
polarizations ($N_p=\pm 1$) is split in two or more peaks. This is
in contrast to the Darwin-Fock energy diagram in Fig.\ \ref{E_IM_df_sq} 
from which we have to conclude that the lowest energy 
center-of-mass mode should be unsplit at $B=0$. Certainly it has to
be kept in mind that the states in Fig.\ \ref{E_IM_df_sq} can not 
be assigned any unique quantum numbers, $n_r$ and $M$. The evolution
from one absorption peak to many at higher deviation is seen in 
Fig.\ \ref{fig_P_IM_sq2_ph}. Observation of the induced density shows
that all the modes are a mixture of center-of-mass  and relative modes
to different extent here. Some peaks can be identified with
depolarization shifted transitions in the interacting Hartree energy
spectrum while others are very close to transitions in the
noninteracting Darwin-Fock diagram reflecting center-of-mass
modes.

\begin{figure} 
\epsfxsize 12cm 
\begin{center}
\epsffile{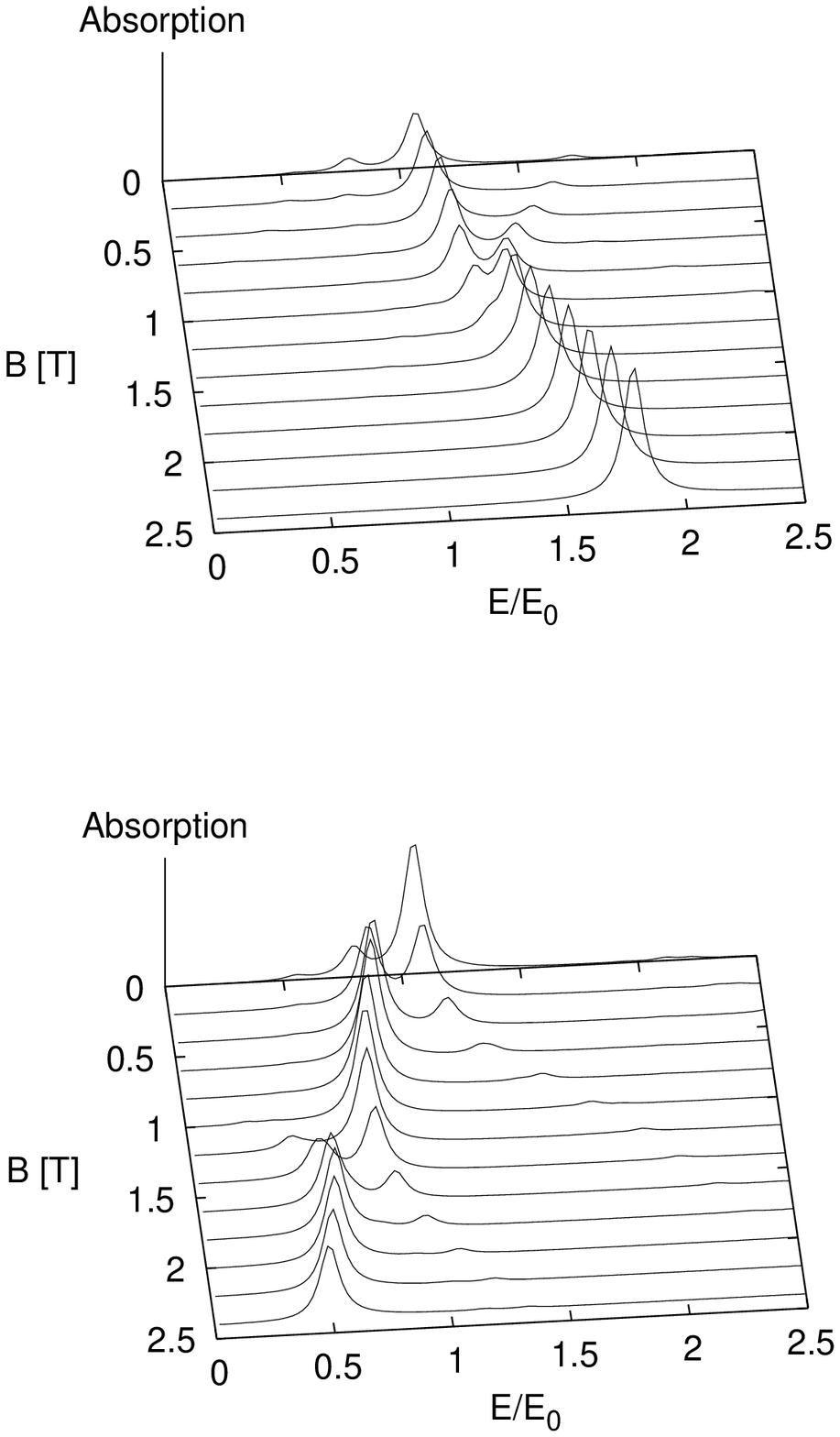}
\end{center}
\caption{The FIR absorption of a square symmetric quantum dot. Same
parameters as in Fig.~\ref{fig_P_IM_sq1} but $N=3$. }
\label{fig_P_IM_sq1_N3}
\end{figure}

\begin{figure} 
\epsfxsize 7.5cm 
\begin{center}
\epsffile{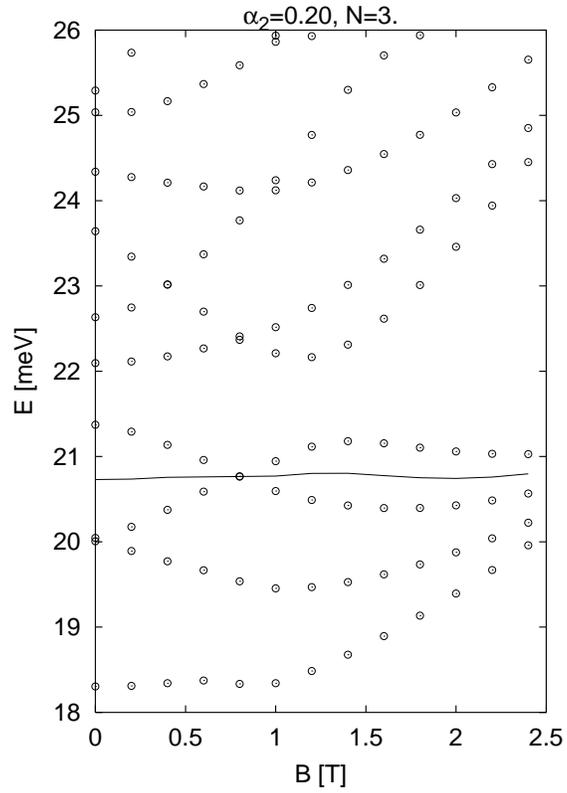}
\end{center}
\caption{The Darwin-Fock energy diagram for $N=3$ interacting
electrons and $\alpha_2=0.20$. The chemical potential is denoted with
the solid line. } 
\label{fig_EH_IM_df_sq}
\end{figure}

In Fig.\ \ref{fig_P_IM_sq1_N3} the FIR absorption is
shown for three electrons, ($N=3$). A splitting 
or anticrossing occurs in the
absorption for $N=3$ and $N_p=-1$ whereas it does not for $N=2$. 
This
striking behavior can be understood from the Darwin-Fock energy
diagram for three electrons (Fig.~\ref{fig_EH_IM_df_sq}). The
splitting can be identified as a transition from the third energy
level to the fourth and fifth with a depolarization shift,
showing its many-body character. This behavior can not be seen for
$N=2$, since the third level is then not occupied.

\begin{figure} 
\epsfxsize 12cm 
\begin{center}
\begin{turn}{-90}
\epsffile{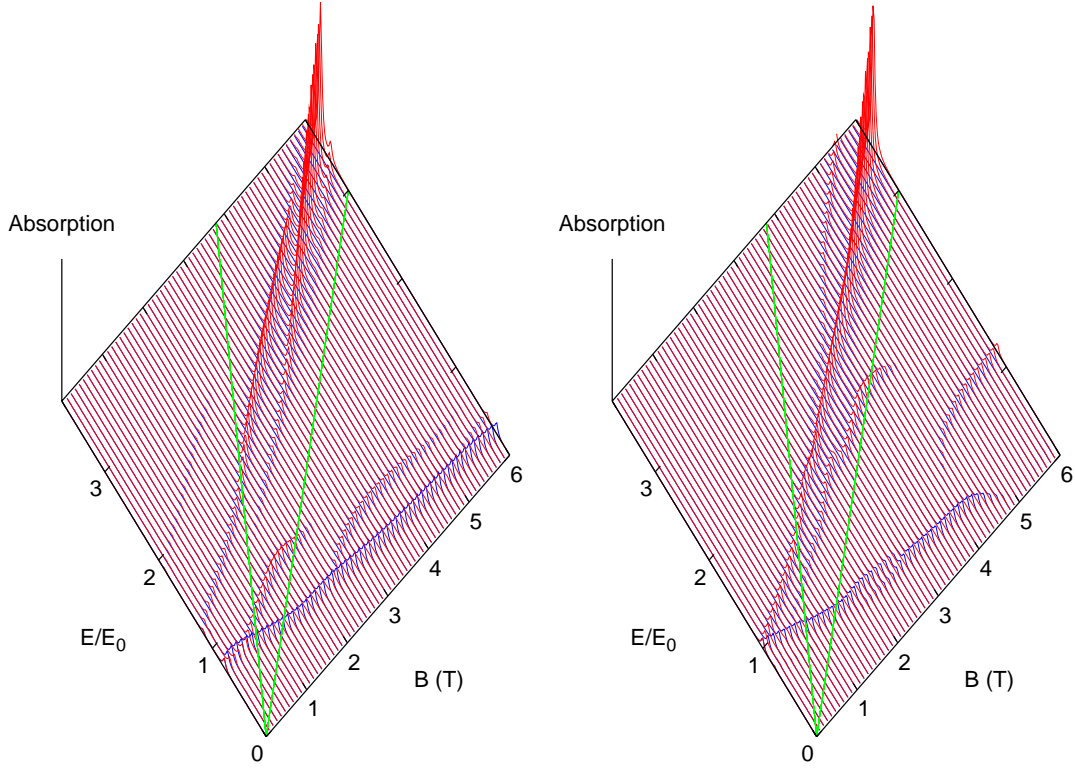}
\end{turn}
\end{center}
\caption{The far-infrared absorption $P(E/E_0,B)$ in the
         case of $N=2$ (left), and $N=12$ (right) electrons
         in a circular symmetric quantum dot with the
         center removed. The straight lines indicate the 
         cyclotron resonance and its first harmonic.
         Only the lowest branch in each case corresponds to $N_p=-1$
         and all other branches belong to $N_p=+1$.
         $E_0=\hbar\omega_0=3.37$ meV, $T=1.0$ K.}
\label{fig_P_vg}
\end{figure}

\begin{figure} 
\epsfxsize 11cm 
\begin{center}
\begin{turn}{-90}
\epsffile{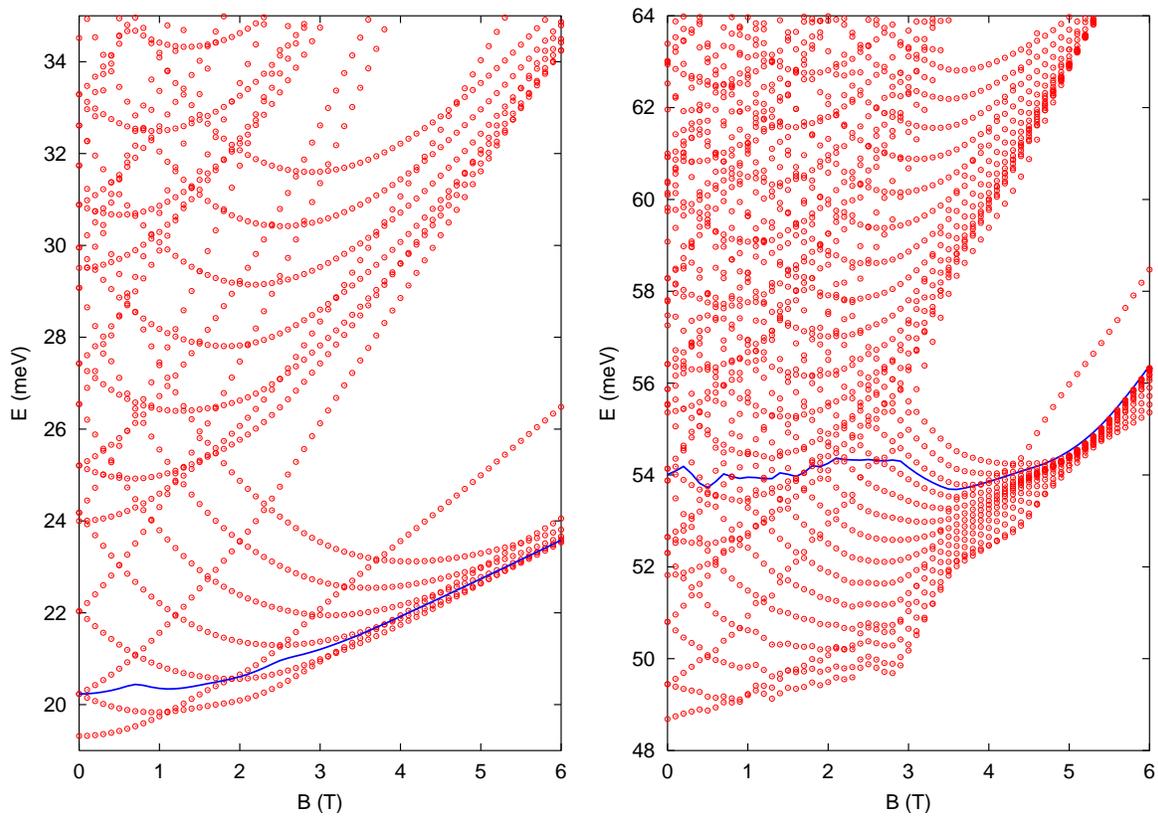}
\end{turn}
\end{center}
\caption{The Darwin-Fock energy diagram in the
         case of $N=2$ (left), and $N=12$ (right) electrons
         in a circular symmetric quantum dot with the
         center removed. The chemical potential $\mu$ is
         indicated by the solid curve.}
\label{fig_E_vg}
\end{figure}

Generally the lower energy branch (seen in $N_p=-1$ polarization
in circular dots) is very stable against splitting, especially 
for dots with many electrons. One exception is a dot with some
kind of a hole through its center. Such dots have been grown as
self-organized InAs rings embedded into a GaAs/AlGaAs heterostructure.   
Here we try to model simply the qualitative behavior of such dots
by using GaAs parameters and by replacing the hole with a repulsive
Coulomb center in the middle of the dot. The confinement of the 
electrons is thus not of the same type as was used to study the
far-infrared absorption of a finite 2DEG with an embedded
impurity.\cite{Gudmundsson94:17433} The calculated absorption
is seen in Fig.\ \ref{fig_P_vg} for two and twelve electrons
in the quantum ring. To understand the absorption one has to
consider the Darwin-Fock energy spectrum of the Hartree interacting
electrons in the ring shown in Fig.\ \ref{fig_E_vg}.
In the case of two electrons in the ring the emptying of states
with low angular quantum number $M$ close to the center with
increasing magnetic field $B$ results in low energy 'side branches`
with polarization $N_p=+1$ to the $N_p=-1$ branch. The magnetic
length, $a$, decreases with increasing $B$, causing the electrons to
'see' in a sence more of the repulsice Coulomb center (and each
other), resulting in the emptying of states of low $M$. It is our
suggestion that A.\ Lorke is observing this same effect in his
experiment.\cite{Lorke98:424} It should be mentioned here that the
electron density in our calculation at the center of the ring
is low for $B=0$ and vanishes as $B$ increases.
In the case of twelve electrons screening effects and 
oscillating structure
reminiscent of emptying of Landau bands is quite clear in the
Darwin-Fock energy diagram, but again emptying of the $M=0$
state at $B=4$ T is visible in the absorption together with
restructuring of the states with high $M$ at the outer edge.     
\section{Summary}
We have used a general two-dimensional multipolar expansion
of the confinement potential of a single quantum dot to
study the effects of shape on its properties.
We have shown that the far-infrared absorption spectrum of
quantum dots with few electrons strongly reflects their shape
and exact number of electrons if the confinement potential
does not allow the application of Kohn's theorem. 
Elliptic dots have two low energy peaks of equal strength
at $B=0$, whereas, dots with slight square symmetry reveal
only one peak. This behavior is connected to the different
effects of shape on the ($M$,$n_r$)$=$($\pm 1$,1) degeneracy
of circular dots at $B=0$. We find that this
degeneracy can be lifted by the activation of relative modes 
in the far-infrared absorption spectrum of square shaped 
dots with increased deviation from the circular form. 

In quantum rings we observe a shift of occupation of the
states labelled by the angular momentum quantum number $M$
to higher values with increasing magnetic field. This is
essentially caused by the changing ratio of the magnetic
length, $a$, and the dots radii. We identify structures 
seen in experiments with this effect.\cite{Lorke98:424}    

The Hartree approximation has been used here rather than the 
more desirable Hartree-Fock approximation in some unrestricted
version, in order to manage the size of the absorption part of
the calculation measured in GBytes and CPU-time. On the other hand,
the HA can be expected to give a valid description of a system
whose modes of excitation are mainly of the center-of-mass type.   

%
%
\acknowledgments
This research was supported by the Icelandic Natural Science Foundation
and the University of Iceland Research Fund. 
%
%

%
%
\bibliographystyle{prsty}

%
%
%
%
%
%
%
%
%
%
%
%
%
%
%
%
%
%
%
%
%
%
%
\end{document}